\pdfoutput=1

\documentclass[11pt,twoside,a4paper,cmspaper,final,collab]{cms-tdr}
\def\svnVersion{464817P}\def\cmsCernNoTag{CERN-EP-2017-240}\def\cmsCernDate{\today}\def\cmsMessage{Published in Physics Letters B as \href{http://dx.doi.org/10.1016/j.physletb.2018.04.030}{\doi{10.1016/j.physletb.2018.04.030.}}}

\begin{document}\cmsNoteHeader{BPH-15-008}

\hyphenation{had-ron-i-za-tion}
\hyphenation{cal-or-i-me-ter}
\hyphenation{de-vices}
\RCS$HeadURL: svn+ssh://svn.cern.ch/reps/tdr2/papers/BPH-15-008/trunk/BPH-15-008.tex $
\RCS$Id: BPH-15-008.tex 452276 2018-03-22 18:22:56Z alverson $
\newlength\cmsFigWidth
\ifthenelse{\boolean{cms@external}}{\setlength\cmsFigWidth{0.85\columnwidth}}{\setlength\cmsFigWidth{0.4\textwidth}}
\ifthenelse{\boolean{cms@external}}{\providecommand{\cmsLeft}{top\xspace}}{\providecommand{\cmsLeft}{left\xspace}}
\ifthenelse{\boolean{cms@external}}{\providecommand{\cmsRight}{bottom\xspace}}{\providecommand{\cmsRight}{right\xspace}}
\ifthenelse{\boolean{cms@external}}{\providecommand{\cmsBottom}{bottom}}{\providecommand{\cmsBottom}{bottom}}
\ifthenelse{\boolean{cms@external}}{\providecommand{\cmsLeft}{top}}{\providecommand{\cmsLeft}{left}}
\ifthenelse{\boolean{cms@external}}{\providecommand{\cmsRight}{bottom}}{\providecommand{\cmsRight}{right}}
\newcommand{\x}{\ensuremath{\phantom{0}}}
\newcommand{\y}{\ensuremath{\phantom{.}}}

\providecommand{\cPKst}{\ensuremath{\cmsSymbolFace{K}^\ast}\xspace}
\providecommand{\cPKstz}{\ensuremath{\cmsSymbolFace{K}^{\ast0}}\xspace}
\providecommand{\cPAKstz}{\ensuremath{\overline{\cmsSymbolFace{K}}{}^{\ast0}}\xspace}
\newcommand{\BtoKstmumu}{\ensuremath{\PBz\to\cPKstz \Pgmp \Pgmm}\xspace}
\newcommand{\BtoKstJpsi}{\ensuremath{\PBz\to\cPJgy \cPKstz}\xspace}
\newcommand{\BtoKstpsip}{\ensuremath{\PBz\to\psi' \cPKstz}\xspace}
\newcommand{\BtoKstJpsimumu}{\ensuremath{\PBz\to\cPJgy(\Pgmp \Pgmm \cPKstz})\xspace}
\newcommand{\BtoKstpsipmumu}{\ensuremath{\PBz\to\psi'(\Pgmp \Pgmm) \cPKstz}\xspace}
\newcommand{\BtoKstmumudecay}{\ensuremath{\PBz\to\cPKstz(\PKp \Pgpm) \Pgmp \Pgmm}\xspace}
\newcommand{\BtoKstJpsidecay}{\ensuremath{\PBz\to\cPJgy(\Pgmp \Pgmm) \cPKstz(\PKp \Pgpm)}\xspace}
\newcommand{\BtoKstpsipdecay}{\ensuremath{\PBz\to\psi'(\Pgmp \Pgmm) \cPKstz(\PKp \Pgpm)}\xspace}
\newcommand{\Kstmumudecay}{\ensuremath{\cPKstz(\PKp \Pgpm) \Pgmp \Pgmm}\xspace}
\newcommand{\KstJpsidecay}{\ensuremath{\cPJgy(\Pgmp \Pgmm) \cPKstz(\PKp \Pgpm)}\xspace}
\newcommand{\Kstpsipdecay}{\ensuremath{\psi'(\Pgmp \Pgmm) \cPKstz(\PKp \Pgpm)}\xspace}
\renewcommand{\PaBz}{\ensuremath{\overline{\cmsSymbolFace{B}}{}^{0}}\xspace}
\cmsNoteHeader{BPH-15-008}

\title{Measurement of angular parameters from the decay $\mathrm{B}^0 \to \mathrm{K}^{*0} \mu^+ \mu^-$ in proton-proton collisions at $\sqrt{s}=8\TeV$}

\date{\today}

\abstract{
Angular distributions of the decay $\mathrm{B}^0 \to \mathrm{K}^{*0} \mu^ +\mu^-$ are studied using a sample of proton-proton collisions at \mbox{$\sqrt{s}=8\TeV$}
collected with the CMS detector at the LHC, corresponding to an integrated luminosity of 20.5\fbinv.
An angular analysis is performed to determine the $P_1$ and $P_5'$ parameters, where the $P_5'$ parameter is of particular interest because of recent measurements
that indicate a potential discrepancy with the standard model predictions. Based on a sample of 1397 signal events, the $P_1$ and $P_5'$ parameters are determined
as a function of the dimuon invariant mass squared. The measurements are in agreement with predictions based on the standard model.}

\hypersetup{%
pdfauthor={CMS Collaboration},
pdftitle={Measurement of angular parameters from the decay B0 to K*0 mu+ mu- from proton-proton collisions at sqrt[s] = 8 TeV},
pdfsubject={CMS},
pdfkeywords={CMS, physics, B0 decays}}

\maketitle
\section{Introduction}
\label{sec:Intro}

Phenomena beyond the standard model (SM) of particle physics can become manifest directly, via
the production of new particles, or indirectly, by modifying the production and decay properties of
SM particles. Analyses of flavor-changing neutral-current decays are particularly sensitive to the
effects of new physics because these decays are highly suppressed in the SM.
An example is the decay \BtoKstmumu, where \cPKstz\ indicates the $\PK^{*0}(892)$ meson, with the charge-conjugate
reaction implied here and elsewhere in this Letter unless otherwise stated. An angular
analysis of this decay as a function of the dimuon invariant mass squared $(q^2)$ allows its properties
to be thoroughly investigated.

The differential decay rate for \BtoKstmumu can be written in terms of $q^2$ and three angular variables
as a combination of spherical harmonics, weighted by $q^2$-dependent angular parameters. These angular
parameters in turn depend upon complex decay amplitudes, which are described by Wilson coefficients in
the relevant effective Hamiltonian~\cite{Altmannshofer:2008dz}. There can be different formulations of
the angular parameters. In this Letter we present measurements of the so-called $P_1$ and $P_5'$
parameters~\cite{Matias:2012xw, Matias06}.

New physics can modify the values of these angular parameters~\cite{Altmannshofer:2008dz,
Melikhov:1998cd, Ali:1999mm, Yan:2000dc, Buchalla:2000sk, Feldmann:2002iw, Hiller:2003js, Kruger:2005ep,
Hovhannisyan:2007pb, Egede:2008uy, Hurth:2008jc, Alok:2009tz, Alok:2010zd, Chang:2010zy, DescotesGenon:2011yn,
Matias:2012xw, DescotesGenon:2012zf} relative to the SM~\cite{Bobeth:2010wg, Bobeth:2011nj,
Bobeth:2012vn, Ali:2006ew, Altmannshofer:2008dz, Altmannshofer:2011gn, Jager:2012uw, Descotes-Genon:2013vna}.
While previous measurements of some of these parameters by the BaBar, Belle, CDF, CMS,
and LHCb experiments were found to be consistent with the SM predictions~\cite{BaBar, Belle, CDF, LHCb, CMS:2011, CMS:2012},
the LHCb Collaboration recently reported a discrepancy larger than 3 standard deviations with respect to the
SM predictions for the $P_5'$ parameter~\cite{LHCbP5p1,LHCbP5p2}, and the Belle Collaboration reported a
discrepancy almost as large~\cite{BelleP5p}.

The new measurements of the $P_1$ and $P_5'$ angular parameters in \BtoKstmumu decays presented in this Letter
are performed using a sample of events collected in proton--proton (pp) collisions at a center-of-mass energy
of 8\TeV with the CMS detector at the CERN LHC.
The data correspond to an integrated luminosity of $20.5\pm0.5\fbinv$~\cite{LUMI}.
The \cPKstz\ meson is reconstructed through its decay to $\PKp\Pgpm$, and the \PBz\ meson by fitting
to a common vertex the tracks from two oppositely charged muon candidates and the tracks from the \cPKstz\ decay.
The values of $P_1$ and $P_5'$ are measured by fitting the distributions of events as a function of three angular
variables: the angle between the \Pgmp\ and the \PBz\ in the dimuon rest frame, the angle between the \PKp\
and the \PBz\ in the \cPKstz\ rest frame, and the angle between the dimuon and the $\PK\pi$ decay
planes in the \PBz\ rest frame.
The measurements are performed in the $q^2$ range from 1 to $19\GeV^2$. Data in the ranges $8.68<q^2<10.09\GeV^2$
and $12.90<q^2<14.18\GeV^2$ correspond to \BtoKstJpsi and \BtoKstpsip decays, respectively, and are used as control
samples, since they have the same final state as the nonresonant decays of interest. Here, $\psi'$ denotes the \Pgy\ meson.

CMS previously exploited the same data set used in this analysis to measure two other angular parameters in the
\BtoKstmumu decay as a function of $q^2$: the forward-backward asymmetry of the muons, $A_\mathrm{FB}$, and the \cPKstz\
longitudinal polarization fraction, $F_\mathrm{L}$, as well as the differential branching fraction, $\rd\mathcal{B}/\rd{}q^2$~\cite{CMS:2012}.
After a simplification of the theoretical decay rate expression, this previous measurement was performed using two
out of the three angular variables.
The analysis was performed with a blinded procedure: the definition of fit strategy
and its validation, as well as background distribution determination have been performed on simulated samples, control
region and signal side bands. The final fit on data has been done at the end of validation.
The analysis presented in this Letter shares with the previous analysis, together
with the data set, the criteria used for selecting signal events, which are reported in Section~\ref{sec:Selection}
for completeness.

\section{The CMS detector}
\label{sec:Detector}

A detailed description of the CMS detector, together with the coordinate system and the standard kinematic
variables, can be found in Ref.~\cite{CMS}. The main detector components used in this analysis are the
silicon tracker and the muon detection systems. The silicon tracker, positioned within a superconducting
solenoid that provides an axial magnetic field of 3.8\unit{T}, consists of three pixel layers and ten strip
layers (four of which have a stereo view) in the barrel region, accompanied by similar pixel and strip detectors
in each endcap region, for a total pseudorapidity coverage of $\abs{\eta}<2.5$.
For tracks with transverse momenta $1 < \pt < 10\GeV$ and $\abs{\eta} < 1.4$, the resolutions are
typically 1.5\% in \pt and 25--90 (45--150)\micron in the transverse (longitudinal) impact
parameter~\cite{TRK-11-001}.  Muons are measured in the range $\abs{\eta}< 2.4$ with
detection planes made using three technologies: drift tubes, cathode strip chambers, and
resistive plate chambers.
The probability for a pion, kaon, or proton to be misidentified as a muon is less than $2.5\times10^{-3}$,
$0.5\times10^{-3}$, and $0.6\times10^{-3}$, respectively, for $\pt > 4\GeV$ and $\abs{\eta} < 2.4$.
The muon identification efficiency is greater than 0.80 (0.98) for $\pt > 3.5\GeV$ and $\abs{\eta} < 1.2$ ($1.2 < \abs{\eta} < 2.4$)~\cite{Chatrchyan:2012xi}.
In addition to the tracker and muon detectors, CMS is equipped with electromagnetic and hadronic calorimeters.

Events are selected using a two-level trigger system~\cite{Khachatryan:2016bia}. The first level consists
of specialized hardware processors that use information from the calorimeters and muon systems to select
events of interest at a rate of around 90\unit{kHz}. A high-level trigger processor farm further decreases
the event rate to less than 1\unit{kHz} before data storage.

\section{Reconstruction, event selection, and efficiency}
\label{sec:Selection}

The criteria used to select the candidate events during data taking (trigger) and after full event
reconstruction (offline) make use of the relatively long lifetime of \PBz\ mesons, which leads them
to decay an average of about 1\unit{mm} from their production point. The trigger uses only muon
information to select events, while the offline selection includes the full reconstruction of all
decay products.

All events used in this analysis were recorded with the same trigger, requiring two identified muons
of opposite charge to form a vertex that is displaced from the pp collision region (beamspot).
Multiple pp collisions in the same or nearby beam crossings (pileup) cause multiple vertices in the same event.
The beamspot position (the most probable collision point) and size (the extent of the luminous region covering
68\% of the collisions in each dimension) were continuously measured through Gaussian fits to reconstructed
pileup vertices as part of the online data quality monitoring.  The trigger required each muon to have $\pt
> 3.5\GeV$, $\abs{\eta}<2.2$, and to pass within 2\,cm of the beam axis.  The dimuon system was
required to have $\pt > 6.9\GeV$, a vertex fit $\chi^2$ probability larger than 10\%, and a
separation of the vertex relative to the beamspot in the transverse plane of at least 3 standard deviations,
where the calculation of the standard deviation includes the calculated uncertainty in the vertex position
and the measured size of the beamspot. In addition, the cosine of the angle in the transverse plane between
the dimuon momentum vector and the vector from the beamspot to the dimuon vertex was required to be greater
than 0.9.

The offline reconstruction requires at least two oppositely charged muons and at least two oppositely
charged hadrons. The muons are required to match those that triggered the event.
The matching is performed by requiring an offline muon to match a trigger-level muon within
$\Delta R = \sqrt{\smash[b]{(\Delta\eta)^2+(\Delta\phi)^2}}<0.1$, where $\Delta\eta$ and $\Delta\phi$ are the
pseudorapidity and azimuthal angle differences, respectively, between the directions of the trigger-level
and offline muons.
Offline muons must, in addition, satisfy general muon identification requirements. For example, the muon
track candidate from the silicon  tracker must match a track segment from the muon detector, the $\chi^2$
per degree of freedom in a global fit to the silicon tracker and muon detector hits must be less than 1.9,
there must be at least six silicon tracker hits, including at least two from the pixel detector, and the
transverse (longitudinal) impact parameter with respect to the beamspot must be less than 3 (30)\cm.
These selection criteria are chosen to optimize the muon identification efficiency as measured using
$\cPJgy \to \Pgmp \Pgmm$ decays in data. The dimuon system at the offline level is required to satisfy the
same requirements as specified above for the trigger level.

The charged hadron candidates are required to fail the muon identification criteria, have $\pt>0.8\GeV$,
and an extrapolated distance $d$ of closest approach to the beamspot in
the transverse plane greater than twice the sum in quadrature of the uncertainty in $d$ and the
beamspot transverse size.
For at least one of the two possible identity assignments---that the positively charged hadron is a kaon
and the negatively charged hadron a pion, or vice versa---the invariant mass of the hadron pair must lie
within 90\MeV of the nominal \cPKstz\ mass~\cite{PDG}.
To remove contamination from $\Pgf\to\PKp\PKm$ decays, we temporarily assign the kaon mass to both charged
hadrons, and then eliminate the candidate if the resulting invariant mass of the hadron pair is less than 1.035\GeV.
The \PBz\ candidates are obtained by fitting
the four charged tracks to a common vertex, and applying a vertex constraint to improve the
resolution of the track parameters.  The \PBz\ candidates must have $\pt>8\GeV$,
$\abs{\eta}<2.2$, vertex fit $\chi^2$ probability larger than 10\%, vertex transverse separation $L$
from the beamspot greater than 12 times the sum in quadrature of the uncertainty in $L$ and the
beamspot transverse size, and $\cos{\alpha_{xy}}>0.9994$, where $\alpha_{xy}$ is the angle in the
transverse plane between the \PBz\ momentum vector and the line-of-flight between the beamspot and
the \PBz\ vertex.  The invariant mass $m$ of the \PBz\ candidate must lie within 280\MeV of
the nominal \PBz\ mass $(m_{\PBz})$~\cite{PDG} for either the $\PKm\Pgpp\Pgmp\Pgmm$ or
$\PKp\Pgpm\Pgmp\Pgmm$ possibility. The selection criteria are optimized using signal event samples
from simulation and background event samples from sideband data in $m$.
The sideband includes both a low- and a high-mass region and is defined by $3\sigma_m < \abs{m-m_{\PBz}} < 280\MeV$,
where $\sigma_m$ is the average mass resolution (${\approx}45$\MeV) obtained from fitting a sum of two
Gaussian functions with a common mean to simulated signal events.
After applying the selection criteria, about 5\% of the events have more than one candidate.
A single candidate is chosen based on the best \PBz\ vertex $\chi^2$ probability.

For each of the selected events, the dimuon invariant mass $q$ and its uncertainty $\sigma_{q}$ are calculated.
We define \BtoKstJpsi and \BtoKstpsip control samples through the requirements $\abs{q - m_{\cPJgy}} < 3\sigma_{q}$
and $\abs{q - m_{\psi'}} < 3\sigma_{q}$, respectively, where $m_{\cPJgy}$ and $m_{\psi'}$ are the nominal
masses~\cite{PDG} of the indicated meson. The average value of $\sigma_{q}$ is about 26\MeV.

The remaining event sample still contains contributions from \BtoKstJpsi and \BtoKstpsip decays,
mainly because of unreconstructed soft photons in the charmonium decay, \ie, $\cPJgy\text{ or } \psi' \to \Pgmp \Pgmm \Pgg$.
These events have a low value of $q$ and fall outside the control sample selection described above.
They also have a low value of $m$ and can be selectively removed
using a combined requirement on $q$ and $m$. For $q<m_{\cPJgy}$ $(q>m_{\cPJgy})$, we require
$\abs{(m-m_{\PBz})-(q-m_{\cPJgy})}>160\: (60)\MeV$. For $q<m_{\psi'}$ $(q>m_{\psi'})$, we require
$\abs{(m-m_{\PBz})-(q-m_{\psi'})}>60\: (30)\MeV$.
Using Monte Carlo (MC) simulation, these requirements were set so that less than 10\% of the
background events originate from the control channels.

To avoid bias, the optimization of the selection criteria, and the fit strategy described below
in Section~\ref{sec:Analysis}, are determined before data in the signal region are examined.
The selection criteria do not depend on the choice of the primary vertex, and their optimization
procedure makes use of both MC simulated signal events generated with the same pileup distribution
as in data, and sideband data. After applying these requirements, 3191 events remain.

The selected four-track vertex is identified as a \PBz\ or $\PaBz$ candidate depending on whether the $\PKp\Pgpm$
or $\PKm\Pgpp$ invariant mass is closest to the nominal \cPKstz\ mass. The fraction of
candidates assigned to the incorrect state is estimated from simulation to be 12--14\%, depending on~$q^2$.

The global efficiency, $\epsilon$, is the product of the acceptance and the combined trigger,
reconstruction, and selection efficiencies, all of which are obtained from MC
simulated event samples. The pp collisions are simulated using the \PYTHIA~\cite{Pythia} event generator, version 6.424,
with particle decays described by the \EVTGEN~\cite{EvtGen} generator, version 9.1,
in which final-state radiation is generated using \PHOTOS~\cite{PHOTOS}.
The default matrix element in \PYTHIA is used to describe the events. The simulated particles are propagated through
a detailed model of the detector based on \GEANTfour~\cite{Geant4}.
The reconstruction and selection of the generated events proceed as for the data.
Separate samples of events are generated for \PBz\ decays to $\cPKstz \Pgmp \Pgmm$, $\cPJgy\ \cPKstz$, and $\psi'\ \cPKstz$,
with $\cPKstz \to \PKp \Pgpm$ and both \cPJgy\ and $\psi'$ decaying to $\Pgmp \Pgmm$.
The distribution of pp collision vertices in each sample is adjusted to match the observed distribution.

The acceptance is obtained from generator-level events, \ie, before the particle
propagation with \GEANTfour, and is defined as the fraction of events with
$\pt(\PBz)>8\GeV$ and $\abs{\eta(\PBz)}<2.2$ that satisfy the single-muon
requirements $\pt(\mu)>3.3\GeV$ and $\abs{\eta(\mu)}<2.3$. These criteria
are less restrictive than the final selection criteria in order to account for
finite detector resolution, since they are applied to generator-level quantities.
Only events satisfying the acceptance criteria are processed through the \GEANTfour
simulation, the trigger simulation, and the reconstruction software.

The combined trigger, reconstruction, and selection efficiency is given by
the ratio of the number of events that satisfy the trigger and selection requirements and have a
reconstructed \PBz\ candidate compatible with a generated \PBz\ meson, relative to the number of events that satisfy
the acceptance criteria.
The generated and reconstructed \PBz\ are considered to be compatible if the reconstructed \PKp\ candidate appears
within a distance $\Delta R$ of the generated \PKp\ meson, and analogously for the \Pgpm, \Pgmp, and \Pgmm,
where $\Delta R = 0.3$ for the hadrons and $\Delta R = 0.004$ for the muons. Requiring all four particles in the \PBz\
decay to be matched results in an efficiency of 99.6\% (0.4\% of the events have a correctly reconstructed
\PBz\ candidate that is not matched to a generated \PBz\ meson) and a purity of 99.5\% (0.5\% of the matched
candidates do not correspond to a correctly reconstructed \PBz candidate).
Efficiencies are determined for both correctly tagged (the \PK\ and \Pgp\ have the correct charge)
and mistagged (the \PK\ and \Pgp\ charges are reversed) candidates.

\subsection{Background studies}
\label{ssec:Background}

Using simulation, we search for possible backgrounds that might peak in the \PBz\ mass region.
The event selection is applied to inclusive MC samples of \PBz, \PBs, \PBp, and $\Lambda_{\PQb}$
decays to \cPJgy\ X and $\psi'$ X, where X denotes all of the exclusive decay channels found in
the PDG~\cite{PDG}, and with the \cPJgy\ and $\psi'$ decaying to \Pgmp \Pgmm.
No evidence for a peaking structure near the \PBz\ mass is found. The distributions of the few events
that satisfy the selection criteria are similar to the shape of the combinatorial background.
As an additional check, we generate events with $\PBs \to \cPKstz(\PKp \PGpm) \Pgmp \Pgmm$ decays.
Assuming that the ratio of branching fractions $\mathcal{B}(\PBs \to \cPJgy \cPKstz) / \mathcal{B}(\BtoKstJpsi) \approx 10^{-2}$~\cite{PDG},
less than one event passes our selection criteria.

Possible backgrounds from events with two hadrons misidentified as muons, in particular from the hadronic fully
reconstructable $\PBz \to \PD \mathrm{X}$ decays, are suppressed by the misidentification probability ($10^{-3}\times10^{-3}$),
and are thus considered negligible.
Also, events from $\PBz \to \JPsi \cPKstz$ decays, where a muon and a hadron are swapped, are suppressed by the hadron-to-muon
misidentification probability ($10^{-3}$) and by the muon-to-hadron identification inefficiency
($10^{-1}$). In fact, given the amount of reconstructed $\PBz \to \JPsi \cPKstz$ events (165\,k), we expect ${\approx}16$ events
distributed in the two adjacent $q^2$ bins close to the \cPJgy\ control region.

Backgrounds from semileptonic decays such as $\PBz \to \PDm \PGpp$, $\PBz \to \PDm \PKp$, and $\PBz \to \PDm \Pgmp \PGnGm$,
where \PDm\ decays to $\cPKstz \Pgmm \PAGnGm$ and \cPKstz to $\PKp \PGpm$, are also studied using simulation. We estimate
in data less than one event for each of the three decays populating the low-mass sideband. All these potential sources of
background are evaluated in the whole $q^2$ range, excluding the \cPJgy\ and $\psi'$ control regions, and are found to be negligible.

The impact of other partially reconstructed multibody \PB\ decays that might affect the low-mass sideband is addressed in Section~\ref{sec:Systematics}.

Backgrounds from events in which a $\PBp \to \PKp \Pgmp \Pgmm$ decay is combined with a random pion,
and from events with a $\Lambda_{\PQb} \to \PgL^0 \Pgmp \Pgmm$ decay, where $\PgL^0$ decays to \Pp\ \PK,
in which the proton is assigned the pion mass, are found to be negligible.
In fact, both processes are flavor-changing neutral-current decays, therefore they have a comparable branching fraction to our signal process.
The former decay has a theoretical lower bound on the invariant mass that lies at $\approx$5.41\GeV. We search in data for an invariant
mass peak around the \PBp\ world-average mass after computing the invariant mass for both $\PKp \Pgmp \Pgmm$ and $\PKm \Pgmp \Pgmm$
possibilities in events with $5.41-\sigma_m < m < m_{\PBz} + 0.280\GeV$, but no evidence of such a peak is found.
For the latter decay we search in data for an invariant mass peak around the $\Lambda_{\PQb}$ world-average mass after assigning
the proton mass to the track previously identified as a pion. Also in this case, no evidence of a peak is found.
Indeed, the simulation shows that less than one event is expected to pass our selection requirements.

\section{Analysis method}
\label{sec:Analysis}

\begin{figure*}[t]
\centering
\includegraphics[width=0.99\textwidth]{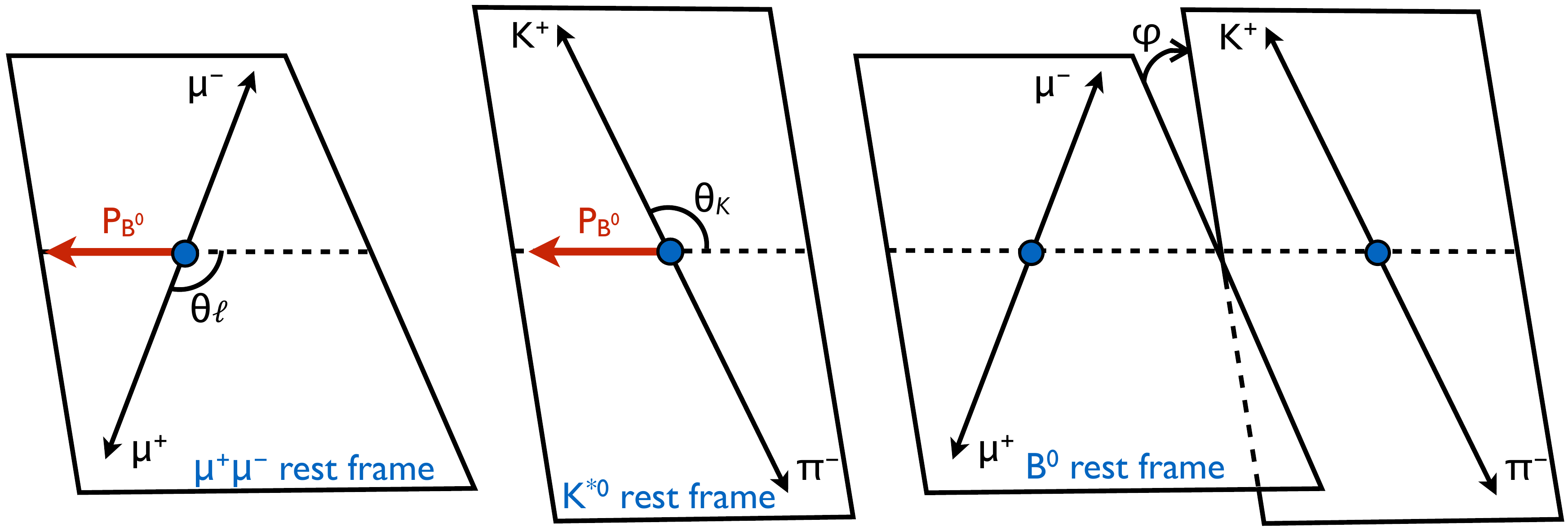}
\caption{Illustration of the angular variables $\theta_{\ell}$ (left), $\theta_{\PK}$ (middle), and $\varphi$ (right) for the decay \BtoKstmumudecay.}
\label{fig:ske}
\end{figure*}

This analysis measures the $P_1$ and $P_5'$ values in \BtoKstmumu decays as a function of $q^2$.
Figure~\ref{fig:ske} illustrates the angular variables needed to describe the decay: $\theta_{\ell}$
is the angle between the positive (negative) muon momentum and the direction opposite to the
\PBz\ $\big(\PaBz\big)$ momentum in the dimuon rest frame,
$\theta_{\PK}$ is the angle between the kaon momentum and the direction opposite to
the \PBz\ $\big(\PaBz\big)$ momentum in the \cPKstz\ $\big(\cPAKstz\big)$ rest frame,
and $\varphi$ is the angle between the plane containing the two muons and the plane containing the
kaon and the pion in the \PBz\ rest frame.
Although the $\PKp\Pgpm$ invariant mass is required to be consistent with that of a \cPKstz\ meson,
there can be a contribution from spinless (S-wave) $\PKp\Pgpm$
combinations~\cite{Descotes-Genon:2013vna,Becirevic:2012dp,Matias:2012qz,Blake:Swave}. This is
parametrized with three terms: $F_{\mathrm{S}}$, which is related to the S-wave fraction, and
$A_{\mathrm{S}}$ and $A^5_{\mathrm{S}}$, which are the interference amplitudes between the S- and
P-wave decays. Including these components, the angular distribution of \BtoKstmumu decays can be
written as~\cite{Descotes-Genon:2013vna}:
\begin{equation} \label{eq:PDF}
\ifthenelse{\boolean{cms@external}}
{
\begin{split}
\frac{1} {\rd \Gamma / \rd{}q^2} &  \frac{\rd^4\Gamma} {\rd{}q^2\ \rd\!\cos\theta_{\ell}\ \rd\!\cos\theta_{\PK}\ \rd\varphi} = \\
\frac{9} {8\pi}  \left\{ \ \frac{2}{3}  \right. & \left[ \ (F_{\mathrm{S}} + A_{\mathrm{S}}\cos\theta_{\PK}) \left( 1-\cos^2\theta_{\ell} \right)  \right.  \\  & \left. + \ A^5_{\mathrm{S}} \sqrt{1-\cos^2\theta_{\PK}}  \sqrt{1-\cos^2\theta_{\ell}}\cos\varphi \ \right] \\
 +  \left(1 - F_{\mathrm{S}} \right) \  & \Bigl[ \ 2\,F_{\mathrm{L}}\cos^2\theta_{\PK} \left( 1-\cos^2\theta_{\ell} \right) \Bigr. \\
& + \frac{1} {2} \left( 1-F_{\mathrm{L}} \right) \left( 1-\cos^2\theta_{\PK} \right) \left( 1+\cos^2\theta_{\ell} \right) \\ & + \frac{1} {2} P_1(1-F_{\mathrm{L}}) \\ & \ \ \ \ \
(1-\cos^2\theta_{\PK})(1-\cos^2\theta_{\ell})\cos2 \varphi \\ & + 2\,P_5'\cos\theta_{\PK} \sqrt{F_{\mathrm{L}} \left( 1-F_{\mathrm{L}} \right) } \\ & \ \ \ \ \Biggl. \Bigl.  \sqrt{1-\cos^2\theta_{\PK}} \sqrt{1-\cos^2\theta_{\ell}}\cos\varphi \ \Bigr] \ \Biggr\},
\end{split}
}
{
\begin{split}
\frac{1} {\rd \Gamma / \rd{}q^2} \frac{\rd^4\Gamma} {\rd{}q^2\ \rd\!\cos\theta_{\ell}\ \rd\!\cos\theta_{\PK}\ \rd\varphi} & =
 \frac{9} {8\pi} \left\{ \frac{2}{3}\, \biggl[ \ (F_{\mathrm{S}}  + A_{\mathrm{S}}\cos\theta_{\PK}) \left( 1-\cos^2\theta_{\ell} \right) \biggr. \right.  \\  & \biggl. \quad \quad + \ A^5_{\mathrm{S}} \,\sqrt{1-\cos^2\theta_{\PK}} \sqrt{1-\cos^2\theta_{\ell}}\cos\varphi \ \biggr] \\  +  \left(1 - F_{\mathrm{S}} \right)  \Bigl[ \Bigr.  & \ \ 2\, F_{\mathrm{L}}\cos^2\theta_{\PK} \left( 1-\cos^2\theta_{\ell} \right) \\
 + & \  \frac{1} {2} \left( 1-F_{\mathrm{L}} \right) \left( 1-\cos^2\theta_{\PK} \right) \left( 1+\cos^2\theta_{\ell} \right) \\
 + & \  \frac{1} {2}\, P_1\, (1-F_{\mathrm{L}}) (1-\cos^2\theta_{\PK})(1-\cos^2\theta_{\ell})\cos2 \varphi \\
 + & \  \bigl. \left. 2\, P_5'\, \cos\theta_{\PK} \sqrt{F_{\mathrm{L}} \left( 1-F_{\mathrm{L}} \right) } \sqrt{1-\cos^2\theta_{\PK}} \sqrt{1-\cos^2\theta_{\ell}}\cos\varphi \  \Bigr] \right\},
\end{split}
}
\end{equation}
where $F_{\mathrm{L}}$ denotes the longitudinal polarization fraction of the \cPKstz.
This expression is an exact simplification of the full angular distribution, obtained by folding the $\varphi$ and
$\theta_{\ell}$ angles about zero and $\pi/2$, respectively.
Specifically, if $\varphi < 0$, then $\varphi \to -\varphi$, and the new $\varphi$ domain is [0, $\pi$]. If $\theta_{\ell} > \pi/2$, then
$\theta_{\ell} \to \pi - \theta_{\ell}$, and the new $\theta_{\ell}$ domain is [0, $\pi/2$].
We use this simplified version of the expression because of difficulties in the fit convergence with the full angular distribution
due to the limited size of the data sample. This simplification exploits the odd symmetry of the angular variables with respect
to $\varphi = 0$ and $\theta_{\ell} = \pi/2$ in such a manner that the cancellation around these angular values is exact.
This cancellation remains approximately valid even after accounting for the experimental acceptance because the efficiency is symmetric
with respect to the folding angles.

For each $q^2$ bin, the observables of interest are extracted from an unbinned extended maximum-likelihood fit
to four variables: the $\PKp\Pgpm\Pgmp\Pgmm$ invariant mass $m$ and the three angular variables ${\theta_{\ell}}$,
${\theta_{\PK}}$, and $\varphi$.
The unnormalized probability density function (pdf) in each $q^2$ bin has the following form:
\begin{equation} \label{eq:angALL}
\ifthenelse{\boolean{cms@external}}
{
\begin{split}
\mathrm{pdf}& (m,\theta_{\PK},\theta_{\ell},\varphi)  = \\  Y^{C}_{S}\, & \biggl[\, S^{C}(m)  \, S^a(\theta_{\PK},\theta_{\ell},\varphi) \, \epsilon^{C}(\theta_{\PK},\theta_{\ell},\varphi) \biggr. \\
+ \ & \biggl.  \frac{f^{M}}{1-f^{M}}~S^{M}(m) \, S^a(-\theta_{\PK},-\theta_{\ell},\varphi) \, \epsilon^{M}(\theta_{\PK},\theta_{\ell},\varphi) \, \biggr] \\
   &  + Y_{B}\,B^m(m) \, B^{\theta_{\PK}}(\theta_{\PK}) \, B^{\theta_{\ell}}(\theta_{\ell}) \, B^{\varphi}(\varphi), \\
\end{split}
}
{
\begin{split}
\mathrm{pdf}(m,\theta_{\PK},\theta_{\ell},\varphi) & = Y^{C}_{S} \biggl[ S^{C}(m)  \, S^a(\theta_{\PK},\theta_{\ell},\varphi) \, \epsilon^{C}(\theta_{\PK},\theta_{\ell},\varphi) \biggr. \\
& \biggl. + \frac{f^{M}}{1-f^{M}}~S^{M}(m) \, S^a(-\theta_{\PK},-\theta_{\ell},\varphi) \, \epsilon^{M}(\theta_{\PK},\theta_{\ell},\varphi) \biggr] \\
& + Y_{B}\,B^m(m) \, B^{\theta_{\PK}}(\theta_{\PK}) \, B^{\theta_{\ell}}(\theta_{\ell}) \, B^{\varphi}(\varphi), \\
\end{split}
}
\end{equation}
where the three terms on the righthand side correspond to correctly tagged signal events, mistagged signal events,
and background events. The parameters $Y^{C}_{S}$ and $Y_{B}$ are the yields of correctly tagged signal events and
background events, respectively, and are determined in the fit. The parameter $f^{M}$ is the fraction of signal events
that are mistagged and is determined from simulation. Its value ranges from 0.124 to 0.137 depending on the $q^2$ bin.

The signal mass probability functions $S^{C}(m)$ and $S^{M}(m)$ are each the sum of two Gaussian functions,
with a common mean for all four Gaussian functions, and describe the mass distribution for correctly tagged and
mistagged signal events, respectively.
In the fit, the mean, the four Gaussian function's width parameters, and the two fractions specifying the relative
contribution of the two Gaussian functions in $S^{C}(m)$ and $S^{M}(m)$ are determined from simulation.
The function $S^a(\theta_{\PK},\theta_{\ell},\varphi)$ describes the signal in the three-dimensional (3D) space of the
angular variables and corresponds to Eq.~(\ref{eq:PDF}).
The combination $B^m(m) \, B^{\theta_{\PK}}(\theta_{\PK}) \, B^{\theta_{\ell}}(\theta_{\ell}) \, B^{\varphi}(\varphi)$ is obtained
from the \PBz\ sideband data in $m$ and describes the background in the space of $(m,\theta_{\PK},\theta_{\ell},\varphi)$,
where $B^m(m)$ is an exponential function, $B^{\theta_{\PK}}(\theta_{\PK})$ and $B^{\theta_{\ell}}(\theta_{\ell})$ are second-
to fourth-order polynomials, depending on the $q^2$ bin, and $B^{\varphi}(\varphi)$ is a first-order polynomial.
The factorization assumption of the background pdf in Eq.~(\ref{eq:angALL}) is validated by dividing
the range of an angular variable into two at its center point and comparing the distributions of events
from the two halves in the other angular variables.

The functions $\epsilon^{C}(\theta_{\PK},\theta_{\ell},\varphi)$ and $\epsilon^{M}(\theta_{\PK},\theta_{\ell},\varphi)$ are
the efficiencies in the 3D space of $|\cos\theta_{\PK}| \leq 1$, $0 \leq \cos\theta_{\ell}\leq 1$, and $0\leq \varphi\leq \pi$
for correctly tagged and mistagged signal events, respectively.
The numerator and denominator of the efficiency are separately described with a nonparametric technique, which is
implemented with a kernel density estimator~\cite{opac-b1089297,Cranmer:2000du}. The final efficiency distributions
used in the fit are obtained from the ratio of 3D histograms derived from the sampling of the kernel density estimators.
The histograms have 40 bins in each dimension.
A consistency check of the procedure used to determine the efficiency is performed by dividing
the simulated data sample into two independent subsets, and extracting the angular parameters from the first subset using
the efficiency computed from the second subset. The efficiencies for both correctly tagged and mistagged events peak at
$\cos\theta_{\ell} \approx 0$, around which they are rather symmetric for $q^2<10\GeV^2$, and are approximately flat in
$\varphi$.
The efficiency for correctly tagged events becomes relatively flat in $\cos\theta_{\ell}$ for larger values of $q^2$,
while it has a monotonic decrease for increasing $\cos\theta_{\PK}$ values for $q^2<14\GeV^2$.
For larger values of $q^2$ a decrease in the efficiency is also seen near $\cos\theta_{\PK} = -1$.
The efficiency for mistagged events has a minimum at $\cos\theta_{\ell} \approx 0$ for $q^2>10\GeV^2$,
while it is maximal near $\cos\theta_{\PK} = 0$ for $q^2<10\GeV^2$. For large values of $q^2$
a mild maximum also appears near $\cos\theta_{\PK}=1$.

The fit is performed in two steps.
The initial fit does not include a signal component and uses the sideband data in $m$ to obtain
the $B^m(m)$, $B^{\theta_{\PK}}(\theta_{\PK})$, $B^{\theta_{\ell}}(\theta_{\ell})$, and $B^{\varphi}(\varphi)$
distributions. The distributions obtained in this step are then fixed for the second step, which
is a fit to the data over the full mass range. The fitted parameters in the second step are the
angular parameters $P_1$, $P_5'$, and $A^5_{\mathrm{S}}$, and the yields $Y^{C}_{S}$ and $Y_{B}$.
To avoid difficulties in the convergence of the fit related to the limited number of events,
the angular parameters $F_{\mathrm{L}}$, $F_{\mathrm{S}}$, and $A_{\mathrm{S}}$ are fixed to previous
measurements~\cite{CMS:2012}.

The expression describing the angular distribution of \BtoKstmumu decays, Eq.~(\ref{eq:PDF}), and also its more
general form in Ref.~\cite{Descotes-Genon:2013vna}, can become negative for certain values of the angular parameters.
In particular, the pdf in Eq.~(\ref{eq:angALL}) is only guaranteed to be positive for a particular subset of the
$P_1$, $P_5'$, and $A^5_{\mathrm{S}}$ parameter space. The presence of such a boundary greatly complicates the
numerical maximization process of the likelihood by \textsc{minuit}~\cite{Minuit} and especially the error determination
by \textsc{minos}~\cite{Minuit}, in particular near the boundary between physical and unphysical regions.
Therefore, the second fit step is performed by discretizing the $P_1$, $P_5'$ two-dimensional space and by maximizing the
likelihood as a function of the nuisance parameters $Y^{C}_{S}$, $Y_{B}$, and $A^5_{\mathrm{S}}$ at fixed values of $P_1$ and
$P_5'$. Finally, the distribution of the likelihood values is fit with a bivariate Gaussian distribution.
The position of the maximum of this distribution inside the physical region provides the measurements of $P_1$ and $P_5'$.

The interference terms $A_{\mathrm{S}}$ and $A^5_{\mathrm{S}}$ must vanish if either of the two interfering components vanish.
These constraints are implemented by requiring $\abs{A_{\mathrm{S}}} < \sqrt{\smash[b]{12 F_{\mathrm{S}}(1-F_{\mathrm{S}})F_{\mathrm{L}}}} \, f$
and $\abs{A^5_{\mathrm{S}}} < \sqrt{\smash[b]{3 F_{\mathrm{S}} (1-F_{\mathrm{S}}) (1-F_{\mathrm{L}}) (1+P_1)} }\, f$, where $f$ is a ratio related
to the S- and P-wave line shapes, calculated to be 0.89 near the \cPKstz\ meson mass~\cite{Descotes-Genon:2013vna}.
The constraint on $A_{\mathrm{S}}$ is naturally satisfied since $F_{\mathrm{S}}$, $F_{\mathrm{L}}$, and $A_{\mathrm{S}}$ are
taken from previous measurements~\cite{CMS:2012}.

To ensure correct coverage for the uncertainties, the Feldman--Cousins method~\cite{FC} is used with nuisance parameters.
Two main sets of pseudo-experimental samples are generated.
The first (second) set, used to compute the coverage for $P_1$ ($P_5'$), is generated
by assigning values to the other parameters as obtained by profiling the bivariate Gaussian distribution description of
the likelihood determined from data at fixed $P_1$ ($P_5'$) values. When fitting the pseudo-experimental samples,
the same fit procedure as applied to the data is used.

The fit formalism and results are validated through fits to pseudo-experimental samples,
MC simulation samples, and control channels.  Additional details, including the
size of the systematic uncertainties assigned on the basis of these fits, are described in Section~\ref{sec:Systematics}.

\section{Systematic uncertainties}
\label{sec:Systematics}
The systematic uncertainty studies are described below and summarized in Table~\ref{tab:systematics}
in the same order.

The adequacy of the fit function and the procedure to determine the parameters of interest are validated
in three ways. First, a large, statistically precise MC signal sample with approximately 400 times
the number of events as the data is used to verify that the fitting procedure produces results consistent
with the input values to the simulation.
The difference between the input and output values in this check is assigned as a simulation
mismodeling systematic uncertainty.
It is also verified that fitting a sample with only either correctly tagged or mistagged events
yields the correct results.
Second, 200 subsamples are extracted randomly from the large MC signal sample and combined with background
events obtained from the pdf in Eq.~(\ref{eq:angALL}) to mimic independent data sets of similar size to the
data. These are used to estimate a fit bias by comparing the average values of the results obtained
by fitting the 200 samples to the results obtained using the full MC signal sample.
Much of the observed bias is a consequence of the fitted parameters lying close to the boundaries of the
physical region.
Third, 200 pseudo-experiments, each with the same number of events as the data sample, are generated in
each $q^2$ bin using the pdf in Eq.~(\ref{eq:angALL}), with parameters obtained from the fit to the data.
Fits to these 200 samples do not reveal any additional systematic uncertainty.

\begin{table}[htbp]
\centering
\topcaption{\label{tab:systematics}
Systematic uncertainties in $P_1$ and $P_5'$. For each source, the range indicates the variation over the bins in $q^2$.}
\ifthenelse{\boolean{cms@external}}{\resizebox{\columnwidth}{!}}{}
{
\begin{tabular}{lcccc}
Source & $P_1 (\times 10^{-3})$ & $P_5' (\times 10^{-3})$ \\[1pt]
\hline \\[-2ex]
Simulation mismodeling       &   \x1--33   &  10--23  \\[1pt]
Fit bias                     &   \x5--78   &  \x10--120 \\[1pt]
Finite size of simulated samples  &  29--73   &  \x31--110 \\[1pt]
Efficiency                   &  \x17--100  &   \x5--65  \\[1pt]
$\PK\pi$ mistagging          &   \x\x8--110  &   \x6--66  \\[1pt]
Background distribution      &  12--70   &  10--51  \\[1pt]
Mass distribution            &      12   &      19  \\[1pt]
Feed-through background      &   \x4--12   &   \x3--24  \\[1pt]
$F_{\mathrm{L}}$, $F_{\mathrm{S}}$, $A_{\mathrm{S}}$ uncertainty propagation & \x\x0--210 & \x\x0--210 \\[1pt]
Angular resolution           &   \x2--68   & 0.1--12\y  \\[1pt]
\hline
Total                        & 100--230  &  \x70--250 \\[1pt]
\end{tabular}
}
\end{table}

Because the efficiency functions are estimated from a finite number of simulated events, there is a
corresponding statistical uncertainty in the efficiency.
Alternatives to the default efficiency function are obtained by generating 100 new distributions
for the numerator and the denominator of the efficiency ratio based on the default kernel density
estimators as pdfs, and rederiving new kernel density estimators for each trial.
The effect of these different efficiency functions on the final result
is used to estimate the systematic uncertainty.

The efficiency determination is checked by comparing
efficiency-corrected results obtained from the control channels with the corresponding world-average values.
The \BtoKstJpsi control sample contains 165\,000 events, compared with 11\,000 events for the \BtoKstpsip
sample. Because of its greater statistical precision, we rely on the \BtoKstJpsi sample to perform the check
of the efficiency determination for the angular variables. We do this by measuring the longitudinal polarization
fraction $F_{\mathrm{L}}$ in the \BtoKstJpsi decays. We find $F_{\mathrm{L}} = 0.537 \pm 0.002\stat$, compared with the
world-average value $0.571 \pm 0.007\,\text{(stat+syst)}$~\cite{PDG}. The difference of $0.034$ is propagated to
$P_1$ and $P_5'$ by taking the root-mean-square (RMS) of the respective distributions resulting from refitting the
data 200 times, varying $F_{\mathrm{L}}$ within a Gaussian distribution with a standard deviation of 0.034. As a
cross-check that the overall efficiency is not affected by a $q^2$-dependent offset, we measure the ratio of branching
fractions $\mathcal{B}(\BtoKstpsip) / \mathcal{B}(\BtoKstJpsi) = 0.480 \pm 0.008~({\rm stat}) \pm 0.055~({\rm R}_{\psi}^{\mu\mu})$,
by means of efficiency-corrected yields including both correctly and wrongly tagged events (the same central value
is obtained also separately for the two subsets of events),
where R$_{\psi}^{\mu\mu}$ refers to the ratio $\mathcal{B}(\cPJgy \to \Pgmp \Pgmm) / \mathcal{B}(\psi' \to \Pgmp \Pgmm)$
of branching fractions. This is compared to the world-average value $0.484 \pm 0.018({\rm stat}) \pm 0.011({\rm syst}) \pm 0.012({\rm R}_{\psi}^{\rm ee})$~\cite{PDG},
where ${\rm R}_{\psi}^{\rm ee}$ refers to the corresponding ratio of branching fractions to \Pep\Pem.
The two results are seen to agree within the uncertainties.

To evaluate the uncertainty in the mistag fraction $f^M$, we allow this fraction to vary in a fit to the events
in the \BtoKstJpsi control sample. We find $f^{M} = (14.5 \pm 0.5)\%$, compared to the result from simulation
$(13.7\pm0.1)\%$.
The difference of 0.8 is propagated to $P_1$ and $P_5'$ by determining the RMS of the respective
distributions obtained from refitting the data 10 times, varying $f^{M}$ within a Gaussian distribution
with a standard deviation of 0.8.

\begin{figure*}[htbp]
\centering
\includegraphics[width=0.95\textwidth]{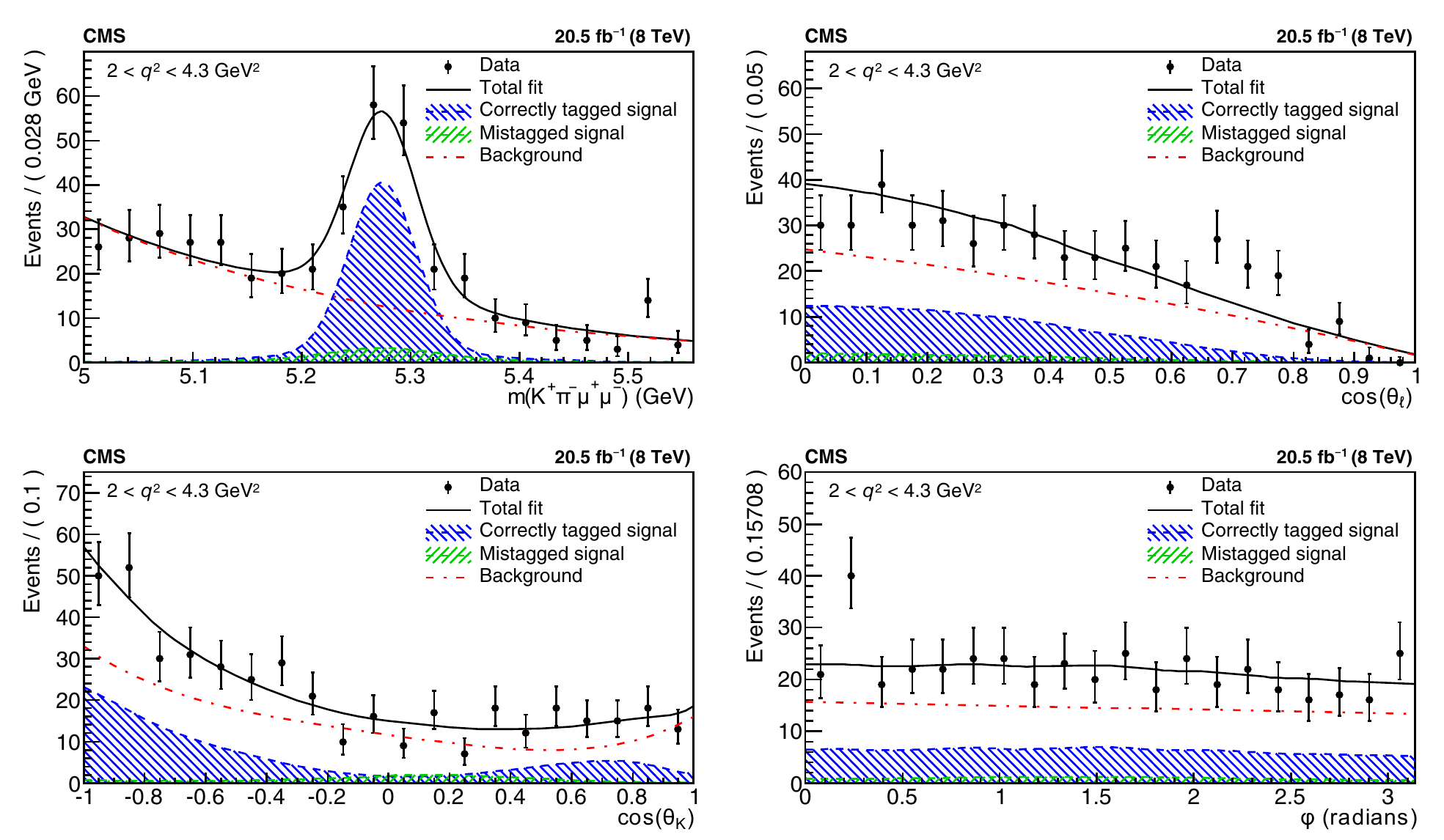}
\includegraphics[width=0.95\textwidth]{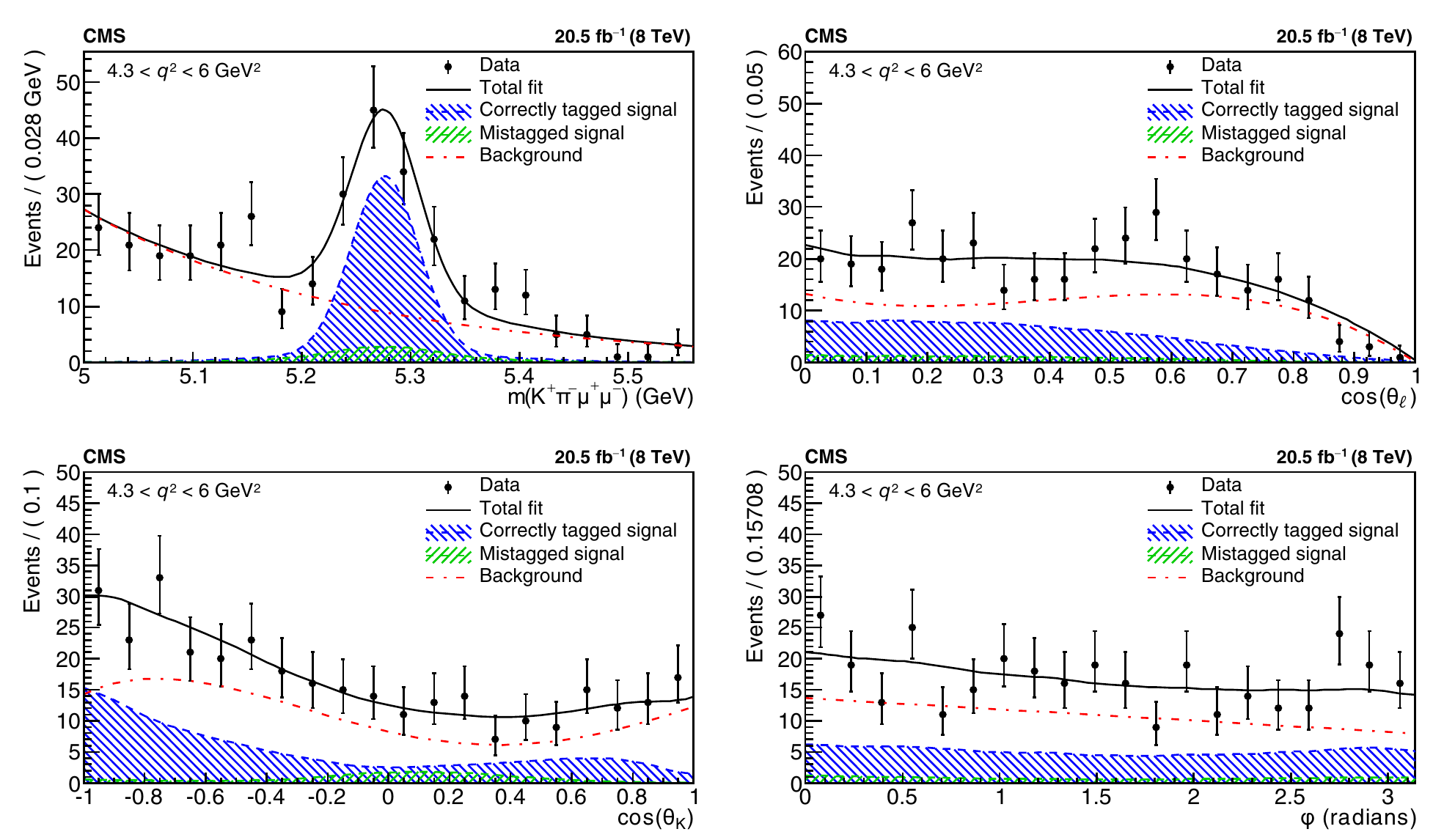}
\caption{Invariant mass and angular distributions of $\PKp\Pgpm\Pgmp\Pgmm$ events for (upper two rows) $2<q^2<4.3\GeV^2$
and (lower two rows) $4.3<q^2<6\GeV^2$. The projection of the results from the total fit, as well as for
correctly tagged signal events, mistagged signal events, and background events, are also shown. The vertical bars
indicate the statistical uncertainties.}
\label{fig:fullplots}
\end{figure*}

The systematic uncertainty associated with the functions used to model the angular distribution of
the background is obtained from the statistical uncertainty in the background shape, as these shapes
are fixed in the final fit.
This uncertainty is determined by fitting the data 200 times, varying the background parameters
within their Gaussian uncertainties, and taking the RMS of the angular parameter values as the
systematic uncertainty. Moreover, for the $q^2$ bin reported in Fig.~\ref{fig:fullplots}, upper two rows, which
shows an excess around $\cos\theta_{\ell} \approx 0.7$ that is also present in the sideband distribution (not
shown in the figure), we refit the data using different descriptions of the background as a function of
$\cos\theta_{\ell}$. The differences in the measurement of $P_1$ and $P_5'$ are within the systematic
uncertainty quoted for the background distribution.

The low-mass sideband might contain partially reconstructed multibody \PBz\ decays.
We test this possibility by refitting the data with a restricted range for the low-mass sideband,
\ie, starting from $\approx$5.1 instead of $\approx$5\GeV.
No significant differences are seen in the measurement of $P_1$ and $P_5'$, and therefore no
systematic uncertainty is assigned.

To evaluate the systematic uncertainty associated with the signal mass pdfs $S^{C}(m)$ and $S^{M}(m)$,
we fit the \BtoKstJpsi and \BtoKstpsip control samples allowing two of the width values
in the four Gaussian terms to vary at a time. The maximum change in $P_1$ and $P_5'$ for either of the
two control channels is taken as the systematic uncertainty for all $q^2$ bins.

The $q^2$ bin just below the \cPJgy\ ($\psi'$) control region, and the $q^2$ bin just above, may be contaminated
with \BtoKstJpsi (\BtoKstpsip) ``feed-through'' events that are not removed by the selection procedure.
A special fit in these two bins is performed, in which an additional background term is added to the pdf.
This background distribution is obtained from simulated \BtoKstJpsi (\BtoKstpsip) events, with the background
yield as a fitted parameter. The resulting changes in $P_1$ and $P_5'$ are used as estimates of the systematic
uncertainty associated with this contribution.

To properly propagate the uncertainty associated with the values of $F_{\mathrm{L}}$, $F_{\mathrm{S}}$, and
$A_{\mathrm{S}}$, taking into account possible correlations, 10 pseudo-experiments per $q^2$ bin are generated
using the pdf parameters determined from the fit to data.
The number of events in these pseudo-experiments is 100 times that of the data.
The pseudo-experiments are then fit twice, once with the same procedure as for the data and once with
$P_1$, $P_5'$, $A^5_{\mathrm{S}}$, $F_{\mathrm{L}}$, $F_{\mathrm{S}}$, and $A_{\mathrm{S}}$
allowed to vary. The average ratio $\rho$ of the statistical uncertainties in
$P_1$ and $P_5'$ from the first fit to that in the second fit is used to compute this systematic uncertainty,
which is proportional to the confidence interval determined from the Feldman--Cousins method through the
coefficient $\sqrt{\smash[b]{\rho^2-1}}$. The stability of $\rho$ as a function of the number of events of
the pseudo-experiments is also verified.
As cross-checks of our procedure concerning the fixed value of $F_{\mathrm{L}}$, we fit the two control regions
either fixing $F_{\mathrm{L}}$ or allowing it to vary, and find that the values of $P_1$ and $P_5'$ are essentially
unaffected, obtaining the same value of $F_{\mathrm{L}}$ as in our previous study~\cite{CMS:2012}. Moreover, we
refit all the $q^2$ bins using only the P-wave contribution for the decay rate in Eq.~(\ref{eq:PDF}) and leaving
all three parameters, $P_1$, $P_5'$, and $F_{\mathrm{L}}$, free to vary.
The differences in the measured values of $P_1$ and $P_5'$ are within the systematic uncertainty quoted for the
$F_{\mathrm{L}}$, $F_{\mathrm{S}}$, and $A_{\mathrm{S}}$ uncertainty propagation.

The effects of angular resolution on the reconstructed values of $\theta_{\PK}$ and $\theta_{\ell}$ are
estimated by performing two fits on the same set of simulated events. One fit uses the true values
of the angular variables and the other fit their reconstructed values. The difference in the
fitted parameters between the two fits is taken as an estimate of the systematic uncertainty.

The systematic uncertainties are determined for each $q^2$ bin, with the total systematic uncertainty
obtained by adding the individual contributions in quadrature.

As a note for future possible global fits of our $P_1$ and $P_5'$ data, the systematic uncertainties
associated with the efficiency, $\PK\pi$ mistagging, \PBz\ mass distribution, and angular resolution can be
assumed to be fully correlated bin-by-bin, while the remaining uncertainties can be assumed to be uncorrelated.

\section{Results}
\label{sec:Results}

\begin{figure*}[htbp!]
\centering
\includegraphics[width=0.49\textwidth]{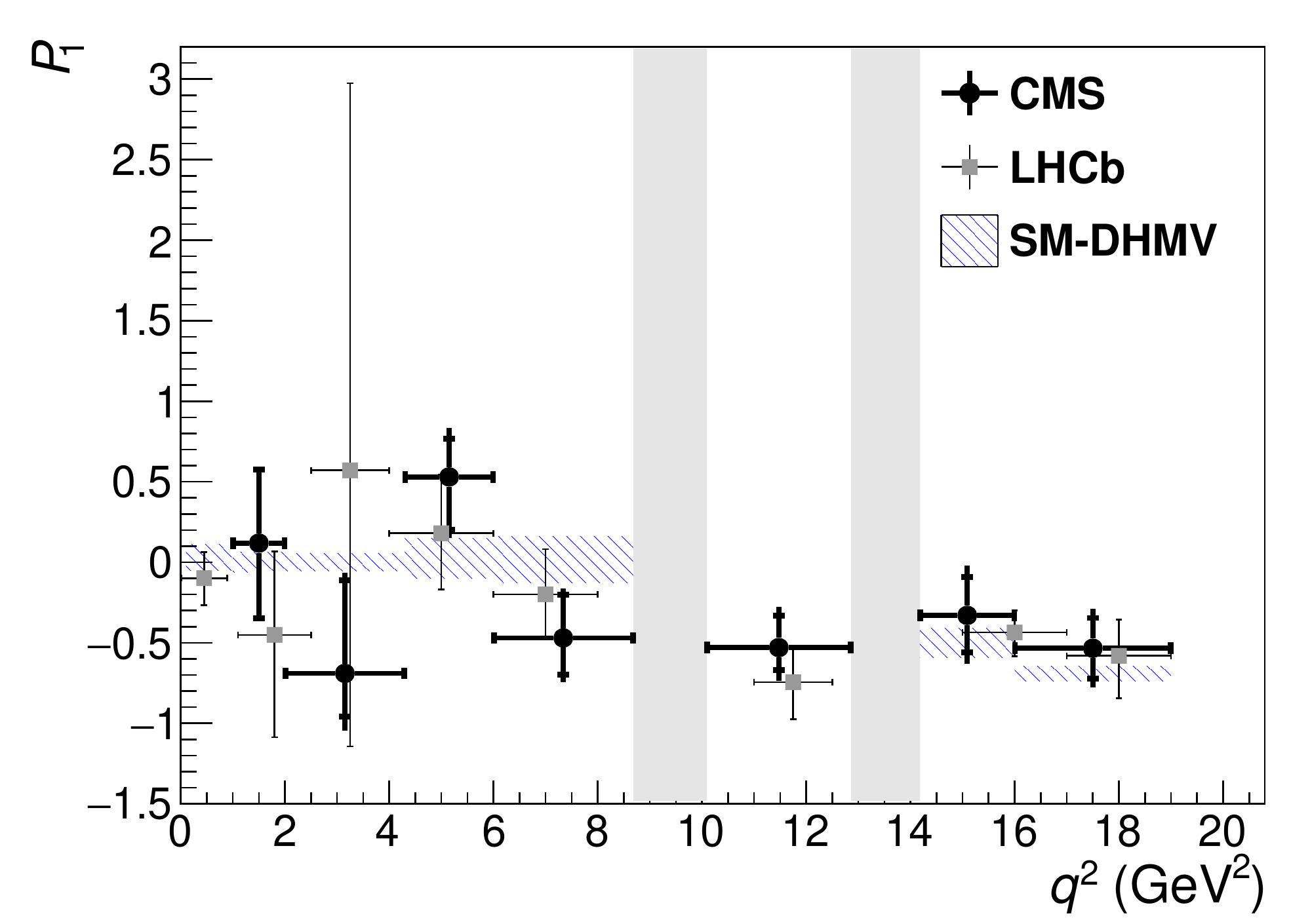}
\includegraphics[width=0.49\textwidth]{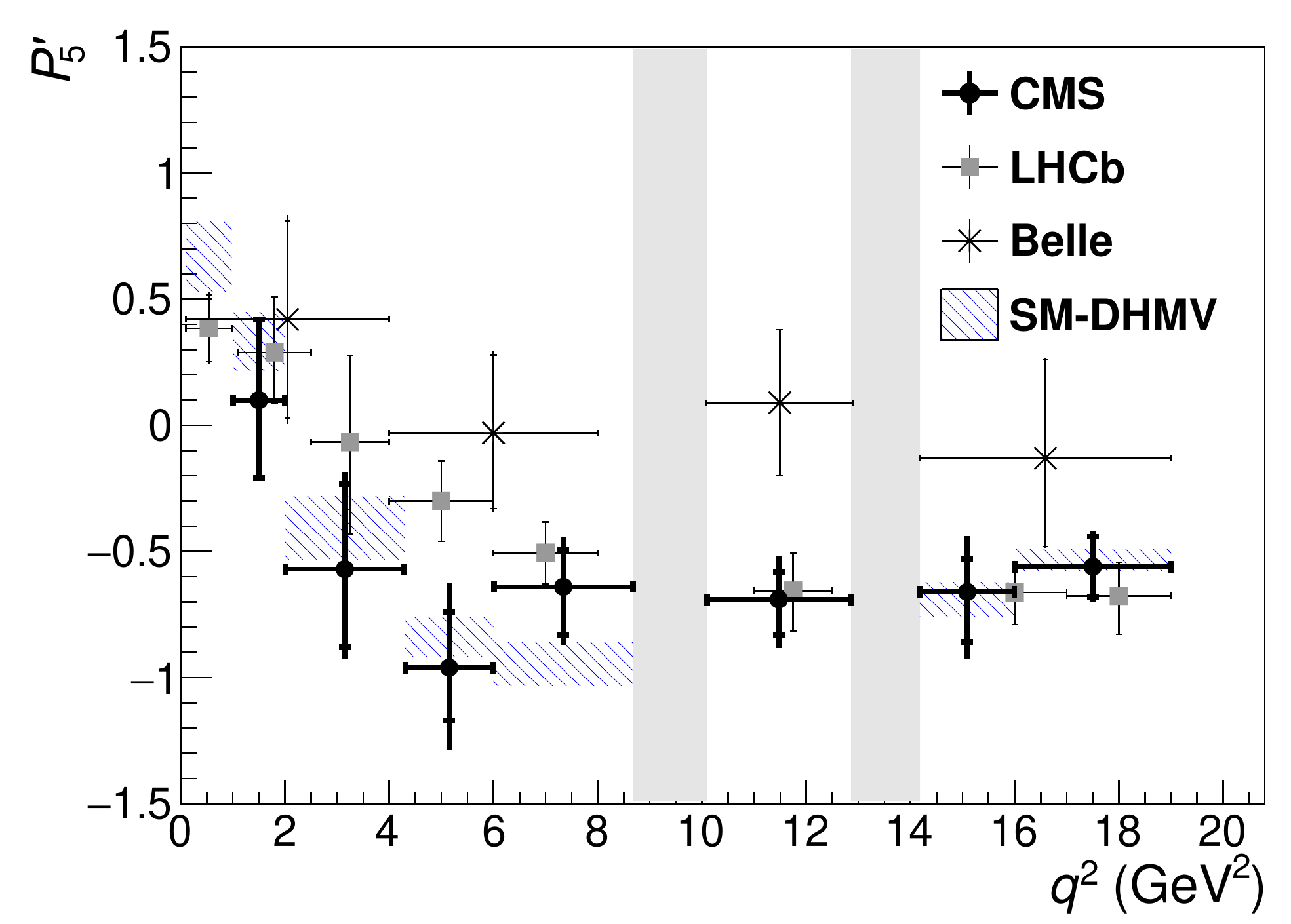}
\caption{CMS measurements of the (left) $P_1$ and (right) $P_5'$ angular parameters versus $q^2$ for \BtoKstmumu decays, in
comparison to results from the LHCb~\cite{LHCbP5p2} and Belle~\cite{BelleP5p} Collaborations.
The statistical uncertainties are shown by the inner vertical bars, while the outer vertical bars give the total uncertainties.
The horizontal bars show the bin widths. The vertical shaded regions correspond to the \cPJgy\ and $\psi'$ resonances.
The hatched region shows the prediction from SM calculations described in the text, averaged over each $q^2$ bin.}
\label{fig:results}
\end{figure*}

The events are fit in seven $q^2$ bins from 1 to $19\GeV^2$, yielding 1397 signal
and 1794 background events in total.
As an example, distributions for two of these bins, along with the fit projections,
are shown in Fig.~\ref{fig:fullplots}.
The fitted values of the signal yields, $P_1$, and $P_5'$ are given in Table~\ref{tab:results}
for the seven $q^2$ bins. The results for $P_1$ and $P_5'$ are shown in Fig.~\ref{fig:results},
along with those from the LHCb~\cite{LHCbP5p2} and Belle~\cite{BelleP5p} experiments.
The fitted values of $A^5_{\mathrm{S}}$ vary from $-0.052$ to $+0.057$.

\begin{table*}[htb]
\centering
\topcaption{\label{tab:results} The measured signal yields, which include both correctly tagged and mistagged events,
the $P_1$ and $P_5'$ values, and the correlation coefficients, in bins of $q^2$, for \BtoKstmumu decays. The first
uncertainty is statistical and the second is systematic. The bin ranges are selected to allow comparison with
previous measurements.}
\begin{tabular}{ccccc}
$q^2~(\GeVns^2)$& Signal yield& $P_1$                            & $P_5'$                          & Correlations \\[1pt]
\hline
1.00--2.00     & $ \x80 \pm 12$ & $+0.12~^{+0.46}_{-0.47} \pm 0.10$ & $+0.10~^{+0.32}_{-0.31} \pm 0.07$ & $-0.0526$ \\[1pt]
2.00--4.30     & $145 \pm 16$ & $-0.69~^{+0.58}_{-0.27} \pm 0.23$ & $-0.57~^{+0.34}_{-0.31} \pm 0.18$ & $-0.0452$ \\[1pt]
4.30--6.00     & $119 \pm 14$ & $+0.53~^{+0.24}_{-0.33} \pm 0.19$ & $-0.96~^{+0.22}_{-0.21} \pm 0.25$ & $+0.4715$ \\[1pt]
6.00--8.68     & $247 \pm 21$ & $-0.47~^{+0.27}_{-0.23} \pm 0.15$ & $-0.64~^{+0.15}_{-0.19} \pm 0.13$ & $+0.0761$ \\[1pt]
10.09--12.86   & $354 \pm 23$ & $-0.53~^{+0.20}_{-0.14} \pm 0.15$ & $-0.69~^{+0.11}_{-0.14} \pm 0.13$ & $+0.6077$ \\[1pt]
14.18--16.00   & $213 \pm 17$ & $-0.33~^{+0.24}_{-0.23} \pm 0.20$ & $-0.66~^{+0.13}_{-0.20} \pm 0.18$ & $+0.4188$ \\[1pt]
16.00--19.00   & $239 \pm 19$ & $-0.53 \pm 0.19         \pm 0.16$ & $-0.56 \pm 0.12         \pm 0.07$ & $+0.4621$ \\[1pt]
\end{tabular}
\end{table*}

A SM prediction, denoted SM-DHMV, is available for comparison with the measured angular parameters.
The SM-DHMV result, derived from Refs.~\cite{DescotesGenon:2012zf,Descotes-Genon:2013vna},
updates the calculations from Ref.~\cite{Ball:2004rg} to account for the known correlation between the different
form factors~\cite{Bharucha:2015}.
It also combines predictions from light-cone sum rules, which are valid in the low-$q^2$ region, with
lattice predictions at high $q^2$~\cite{Horgan:2013hoa} to obtain more precise determinations of the form
factors over the full $q^2$ range. The hadronic charm-quark loop contribution is obtained from Ref.~\cite{Khodjamirian:2010}.
A reliable theoretical prediction is not available near the \cPJgy\ and $\psi'$ resonances.
The SM prediction is shown in comparison to the data in Fig.~\ref{fig:results} and it is seen to be in agreement with the CMS
results. Thus, we do not obtain evidence for physics beyond the SM. Qualitatively, the CMS measurements are compatible with the
LHCb results. The Belle measurements lie systematically above both the CMS and LHCb results and the SM prediction.

\section{Summary}
\label{sec:End}

Using proton-proton collision data recorded at $\sqrt{s}=8\TeV$ with the CMS detector at the LHC, corresponding
to an integrated luminosity of 20.5\fbinv, an angular analysis has been performed for the decay \BtoKstmumu.
The signal sample consists of 1397 selected events. For each of seven bins between 1 to $19\GeV^2$ in the
dimuon invariant mass squared $q^2$, unbinned maximum-likelihood fits are performed on the distributions
of the $\PKp\Pgpm\Pgmp\Pgmm$ invariant mass and three angular variables to obtain values of the $P_1$ and $P_5'$
parameters. The results are among the most precise to date for these parameters and are consistent with
predictions based on the standard model.

\begin{acknowledgments}

We congratulate our colleagues in the CERN accelerator departments for the excellent performance of the
LHC and thank the technical and administrative staffs at CERN and at other CMS institutes for their
contributions to the success of the CMS effort. In addition, we gratefully acknowledge the computing
centres and personnel of the Worldwide LHC Computing Grid for delivering so effectively the computing
infrastructure essential to our analyses. Finally, we acknowledge the enduring support for the
construction and operation of the LHC and the CMS detector provided by the following funding agencies:
BMWFW and FWF (Austria); FNRS and FWO (Belgium); CNPq, CAPES, FAPERJ, and FAPESP (Brazil); MES (Bulgaria);
CERN; CAS, MoST, and NSFC (China); COLCIENCIAS (Colombia); MSES and CSF (Croatia); RPF (Cyprus); SENESCYT
(Ecuador); MoER, ERC IUT, and ERDF (Estonia); Academy of Finland, MEC, and HIP (Finland); CEA and CNRS/IN2P3
(France); BMBF, DFG, and HGF (Germany); GSRT (Greece); OTKA and NIH (Hungary); DAE and DST (India);
IPM (Iran); SFI (Ireland); INFN (Italy); MSIP and NRF (Republic of Korea); LAS (Lithuania); MOE and UM
(Malaysia); BUAP, CINVESTAV, CONACYT, LNS, SEP, and UASLP-FAI (Mexico); MBIE (New Zealand); PAEC (Pakistan);
MSHE and NSC (Poland); FCT (Portugal); JINR (Dubna); MON, RosAtom, RAS, RFBR and RAEP (Russia); MESTD (Serbia);
SEIDI, CPAN, PCTI and FEDER (Spain); Swiss Funding Agencies (Switzerland); MST (Taipei); ThEPCenter, IPST,
STAR, and NSTDA (Thailand); TUBITAK and TAEK (Turkey); NASU and SFFR (Ukraine); STFC (United Kingdom);
DOE and NSF (USA).

\end{acknowledgments}
\vspace*{6ex}

\bibliography{auto_generated}
\cleardoublepage \appendix\section{The CMS Collaboration \label{app:collab}}\begin{sloppypar}\hyphenpenalty=5000\widowpenalty=500\clubpenalty=5000\vskip\cmsinstskip
\textbf{Yerevan Physics Institute,  Yerevan,  Armenia}\\*[0pt]
A.M.~Sirunyan,  A.~Tumasyan
\vskip\cmsinstskip
\textbf{Institut f\"{u}r Hochenergiephysik,  Wien,  Austria}\\*[0pt]
W.~Adam,  F.~Ambrogi,  E.~Asilar,  T.~Bergauer,  J.~Brandstetter,  E.~Brondolin,  M.~Dragicevic,  J.~Er\"{o},  M.~Flechl,  M.~Friedl,  R.~Fr\"{u}hwirth\cmsAuthorMark{1},  V.M.~Ghete,  J.~Grossmann,  J.~Hrubec,  M.~Jeitler\cmsAuthorMark{1},  A.~K\"{o}nig,  N.~Krammer,  I.~Kr\"{a}tschmer,  D.~Liko,  T.~Madlener,  I.~Mikulec,  E.~Pree,  N.~Rad,  H.~Rohringer,  J.~Schieck\cmsAuthorMark{1},  R.~Sch\"{o}fbeck,  M.~Spanring,  D.~Spitzbart,  W.~Waltenberger,  J.~Wittmann,  C.-E.~Wulz\cmsAuthorMark{1},  M.~Zarucki
% ...
\vskip\cmsinstskip
% ...
1:~Also at Vienna University of Technology,  Vienna,  Austria\\
% ...
\end{sloppypar}
\end{document}